\documentclass{aa}  
\usepackage{natbib,twoopt}
\usepackage[breaklinks=true]{hyperref} %% to avoid \citeads line fills
\bibpunct{(}{)}{;}{a}{}{,}             %% natbib format for A&A and ApJ
\makeatletter
  \newcommandtwoopt{\citeads}[3][][]{\href{http://adsabs.harvard.edu/abs/#3}%
    {\def\hyper@linkstart##1##2{}%
     \let\hyper@linkend\@empty\citealp[#1][#2]{#3}}}
  \newcommandtwoopt{\citepads}[3][][]{\href{http://adsabs.harvard.edu/abs/#3}%
    {\def\hyper@linkstart##1##2{}%
     \let\hyper@linkend\@empty\citep[#1][#2]{#3}}}
  \newcommandtwoopt{\citetads}[3][][]{\href{http://adsabs.harvard.edu/abs/#3}%
    {\def\hyper@linkstart##1##2{}%
     \let\hyper@linkend\@empty\citet[#1][#2]{#3}}}
  \newcommandtwoopt{\citeyearads}[3][][]%
    {\href{http://adsabs.harvard.edu/abs/#3}
    {\def\hyper@linkstart##1##2{}%
     \let\hyper@linkend\@empty\citeyear[#1][#2]{#3}}}
\makeatother

\usepackage{graphicx}
\usepackage{txfonts}

\begin{document} 

 \title{Discovery of new globular clusters in the Sagittarius dwarf galaxy}
   \author{
          D. Minniti\inst{1,2}
          \and
          V. Ripepi\inst{3}
          \and
          J. G. Fern\'andez-Trincado\inst{4,5,6}
          \and
          J. Alonso-Garc\'ia\inst{7,8}
          \and
          L. C. Smith\inst{9}
           \and
          P. W. Lucas\inst{10}
          \and
          M. Gómez\inst{1}
          \and
          J. B. Pullen\inst{1}
          \and
          E. R. Garro\inst{1}
          \and
          F. Vivanco C\'adiz\inst{1}
          \and
          M. Hempel\inst{1,11}
          \and
          M. Rejkuba\inst{12}
          \and
          R. K. Saito\inst{13}
         \and
          T. Palma \inst{14}
          \and
          J. J. Clariá \inst{14}
         \and
          M. Gregg \inst{15}
          \and
          D. Majaess \inst{16,17}
          }
   \institute{Departamento de Ciencias Físicas, Facultad de Ciencias Exactas, Universidad Andrés Bello, Fernández Concha 700, Las Condes, Santiago, Chile
\and
 Vatican Observatory, Vatican City State, V-00120, Italy
\and
 INAF-Osservatorio Astronomico di Capodimonte, Salita Moiariello 16, 80131, Naples, Italy
\and
 Instituto de Astronom\'ia y Ciencias Planetarias, Universidad de Atacama, Copayapu 485, Copiap\'o, Chile
\and
Institut Utinam, CNRS UMR 6213, Universit\'e Bourgogne-Franche-Comt\'e, OSU THETA Franche-Comt\'e, Observatoire de Besan\c{c}on, BP 1615, 25010 Besan\c{c}on Cedex, France
\and
Centro de Investigaci\'on en Astronom\'ia, Universidad Bernardo O'Higgins, Avenida Viel 1497, Santiago, Chile
\and
Centro de Astronomía (CITEVA), Universidad de Antofagasta, Av. Angamos 601, Antofagasta, Chile
\and
Millennium Institute of Astrophysics, Santiago, Chile
\and
Institute of Astronomy, University of Cambridge, Madingley Rd, Cambridge CB3 0HA, UK
\and 
Centre for Astrophysics Research, University of Hertfordshire, College Lane, Hatfield, AL10 9AB, UK
\and 
Max-Planck for Astronomy, Koenigstuhl 17, 69117 Heidelberg, Germany.
\and
European Southern Observatory, Karl-Schwarszchild-Str. 2, D85748 Garching b. Munich, Germany
\and 
Departamento de Fisica, Universidade Federal de Santa Catarina, Trindade 88040-900, Florianopolis, SC, Brazil
 \and
 Observatorio Astron\'omico Universidad Nacional de C\'ordoba, Laprida 854, C\'ordoba 5000, Argentina
  \and
 University of California - Davis, Davis CA, USA
 \and
 Department of Chemistry and Physics, Mount Saint Vincent University, Halifax, Nova Scotia B3M 2J6, Canada 
\and
Department of Astronomy and Physics, Saint Mary’s University, Halifax, Nova Scotia B3H 3C3, Canada
}
  \date{Received; Accepted}

% \abstract{}{}{}{}{} 
% 5 {} token are mandatory
 
  \abstract
  % context heading (optional)
  % {} leave it empty if necessary  
   {%CONTEXT. 
Globular clusters (GCs) are witnesses of the past accretion events onto the Milky Way (MW).
In particular, the GCs of the Sagittarius (Sgr) dwarf galaxy are important probes of an on-going merger.}
  % aims heading (mandatory)
   {%AIMS.
   Our main goal is to search for new GC members of this dwarf galaxy using the VISTA Variables in the Via Lactea Extended Survey (VVVX) near-infrared database combined with the Gaia Early Data Release 3 (EDR3) optical database.
   }
  % methods heading (mandatory)
   {%METHODS.
We investigated all VVVX-enabled discoveries of GC candidates in a region covering about 180 sq. deg. toward the bulge and the Sgr dwarf galaxy.  We used multiband point-spread function photometry to obtain deep color-magnitude diagrams (CMDs) and luminosity functions (LFs) for all GC candidates, complemented by accurate Gaia-EDR3 proper motions (PMs) to select Sgr members and variability information to select RR Lyrae which are potential GC  members.
}
  % results heading (mandatory)
   {%RESULTS.
After applying a strict PM cut to discard foreground bulge and disk stars, the CMDs and LFs for some of the GC candidates exhibit well defined red giant branches and red clump giant star peaks. We selected the best Sgr GCs, estimating their distances, reddenings, and associated RR Lyrae.
}
{% CONCLUSION.
We discover 12 new Sgr GC members, more than doubling the number of GCs known in this dwarf galaxy.  
In addition, there are 11 other GC candidates identified that are uncertain, awaiting better data for confirmation.
   }

\titlerunning{New Globular Clusters in the Sagittarius Dwarf Galaxy}
\authorrunning{Dante Minniti, et al.}

   \maketitle
   
\section{Introduction}

Globular clusters (GCs) are important in astrophysics, and one of their latest outstanding applications is that they can be used to characterize possible ancient accretion events that contributed to build the Milky Way (MW) galaxy as we know it. 
By measuring their orbital and chemical properties, the GCs have been used to identify some of these past mergers, such as the Sagittarius (Sgr) dwarf galaxy, Gaia-Enceladus-Sausage, and Sequoia, to mention a few (e.g., Helmi et al. 2018, Belokurov et al. 2018, Massari et al. 2019, Myeong et al. 2019, Barba et al. 2019, Vasiliev 2019, Huang \& Koposov 2020).

It is very important in this context to have a well characterized sample of the GC systems of local dwarf galaxies that could in principle be used as comparisons. There is, however, evidence that the GC systems of the nearby dwarf galaxies may still be incomplete, as shown by the recent discoveries of faint additional GC members of the Fornax, Eridanus II, and Sextans A dwarf spheroidal galaxies (Crnojevic et al. 2016, Wang et al. 2019, Beasley et al. 2019). It is also possible that a large fraction of Local Group dwarf galaxies may have hosted GCs, now in the form of disrupted remnants (Fernandez-Trincado et al. 2020a).

The Sgr dwarf galaxy is  special in this respect. Discovered by Ibata et al. (1994), it exhibits extended tidal tails mapped using different tracers (Mateo et al. 1996, Majewski et al. 2003, Newberg et al. 2003, Correnti et al. 2010, Antoja et a. 2020). 
This merging galaxy contains a few GCs located in its main body, as well as in its extended tails (Majewski et al. 2003, Law \& Majewski 2010). The nine GCs known to be Sgr members are: NGC 6715, which is also the nucleus of this galaxy; Arp 2, Ter , and Ter 8, which are located in the main body of this galaxy; as well as Pal 12, Whiting 1, NGC2419, NGC4147, and NGC5634, which are located in the extended tidal streams (Bellazzini et al. 2020).

However, extensive modeling revealed that this is not a minor merger (e.g., Majewski et al. 2004, Belokurov et al. 2006, Newberg et al. 2007, Hasselquist et al. 2017, 2019, Hayes et al. 2020, Ramos et al. 2020, Vasiliev et al. 2021).
{ The Sgr dwarf is a relatively massive progenitor with a halo mass of $M=10^{11} M_{\odot}$ (e.g., Laporte et al. 2018, Vasiliev \& Belokurov 2020), and
it may be surprising that there are so few star clusters associated with this galaxy. 
However, given the variety of  coverage in terms of GC luminosities from the available studies, we would like to warn readers that before comparing them
with the GC populations of other dwarf galaxies, the correction for faint undetected GCs in these other galaxies has to be taken into account.
Some explanations for a possible GC deficiency may be that: }
(1)  most of the past Sgr GCs were destroyed by dynamical processes (e.g., Fernandez-Trincado et al. 2020b), 
(2)  they were simply not formed because this galaxy quickly ran out of gaseous material, or 
(3)  the GC census of this galaxy is incomplete. 
In particular, GCs are old enough that their orbits are prone to decay onto the central regions of galaxies due to the effect of dynamical friction (e.g., Tremaine \& Weinberg 1984, Gnedin \& Ostriker 1997, Arca-Sedda \& Capuzzo-Dolcetta 2014). 
Dynamical friction is a drag force proportional in strength to the background stellar density. 
This force is expected to bring a globular cluster to the center of its parent galaxy after only a few 
gigayears (e.g., Lotz et al. 2001, Goerdt et al. 2006). 
Therefore, many surviving GCs may still be hidden in the main body of the Sgr galaxy.

{ Additional motivation for this work comes from recent theoretical studies of GC populations in galaxies (Kruijssen et al. 2019, 2020, El Badry et al. 2019, Choski \& Gnedin 2019, Hughes et al. 2019) and from the comparison of GC systems in galaxies with different masses. For example, Kruijssen et al. (2020) assigned $8\pm 3$ massive GCs (with $M> 10^5$ $M_{\odot}$) and a total halo mass $M_T = 10^{10.94} \pm 0.10$ $M_{\odot}$ to the Sgr dwarf galaxy.  This agrees well with the expected number of GCs predicted by Burkert \& Forbes (2020), who find a tight correlation between the number of GCs and the virial mass of the dark matter halos. Their linear relation of $log N_{GC}$ vs $M_{Vir}$ appears to be valid over a large range of galaxy masses ($10^{10}<M_{Vir}< 10^{15})$ $M_{\odot}$ with relatively small scatter (see their Fig. 1). In order to make a more detailed comparison, however, it is useful to have a complete census of the GCs belonging to Sgr. Only in the most nearby galaxies, such as in  Sgr,  can we study the GC systems down to the faintest luminosities.}

The MW GC census is also incomplete in the bulge regions because of large extinction and heavy crowding (Minniti et al. 2017). Following the initial discoveries made a decade ago by the VISTA Variables in the Via Lactea (VVV) survey (Minniti et al. 2011, Moni Bidin et al. 2011, Borissova et al. 2014), over the past years we carried out a systematic large-scale GC search in the inner MW, using the near-IR database of the VVV survey and its extension VVVX (Minniti et al. 2010, Saito et al. 2012), complemented with near-IR data from the Two Micron All Sky Survey (2MASS -- Skutskie et al. 2006), optical data from the DECAM Galactic Plane Survey (DECAPS -- Schlafly et al. 2018), Gaia Data Release Two (DR2 -- The Gaia Collaboration 2018),  the mid-IR data from Glimpse (Benjamin et al. 2003), and the Wide-Field Infrared Survey Explorer (WISE -- Cutri et al. 2012). 

We used different tracers, such as RR Lyrae, type 2 Cepheids, red clump (RC) giants, and blue horizontal branch (BHB) stars, uncovering more than 300 new candidates (e.g., Camargo \& Minniti 2019, Palma et al. 2019, Montenegro et al. 2019, Barba et al. 2019, Minniti et al. 2020 and references therein).
As the Sgr dwarf galaxy is projected behind the Galactic bulge, it is not surprising to expect that some of these new candidate GCs may belong to this galaxy.
We therefore decided to search for GCs in the main body of Sgr and started by exploring the candidate clusters found by the VVV survey and its extension VVVX.
This was  difficult  because of the presence of heavy foreground contamination by disk and bulge field stars along the line of sight.
In fact, this would be impossible without the accurate astrometry provided by the Gaia mission, in particular from the Early Data Release 3 (EDR3 -- Gaia Collaboration, Brown et al. 2021).

In this work, we find 12 new GCs in the Sgr dwarf system, more than doubling the sample of known GCs in this galaxy. 
In addition, we find 11 Sgr objects that are inconclusive and need to be followed up on in order to confirm or discard them as Sgr GCs.
In Section 2 we discuss the data for the new Sgr GCs, including their color-magnitude diagrams (CMDs) and luminosity functions (LFs).
Section 3 presents the RR Lyrae in the GC fields.
Section 4 discusses the physical parameters for the confirmed GCs.
Finally, Section 6 presents the conclusions.

\begin{figure}[h]
\centering
\includegraphics[width=5.4cm, height=9cm]{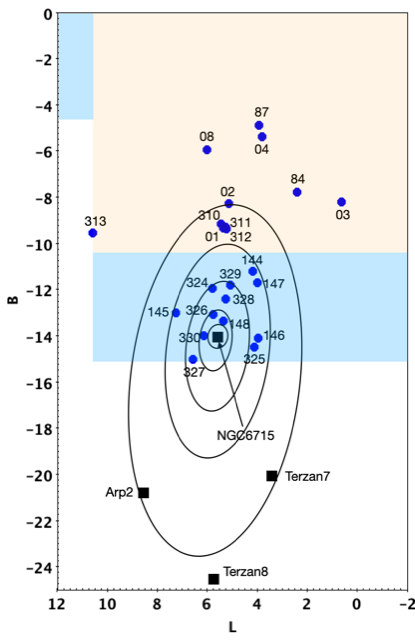} 
\caption{Map showing the location of GCs in Galactocentric coordinates, along with the smoothed density contours from the main body of the Sgr dwarf galaxy (adapted from Bellazzini et al. 2020). Previously known Sgr GCs are shown with black squares and the 23 new GC candidates are shown with blue circles. NGC 6715, the GC nucleus of this galaxy, is marked with the arrow. The areal coverage of the VVV and VVVX near-IR surveys  are depicted in light orange and light blue, respectively.
}
\end{figure}

\section{Data for the new  Sgr GCs} 

\begin{figure*}[h]
\centering
\includegraphics[width=13.2cm, height=10cm]{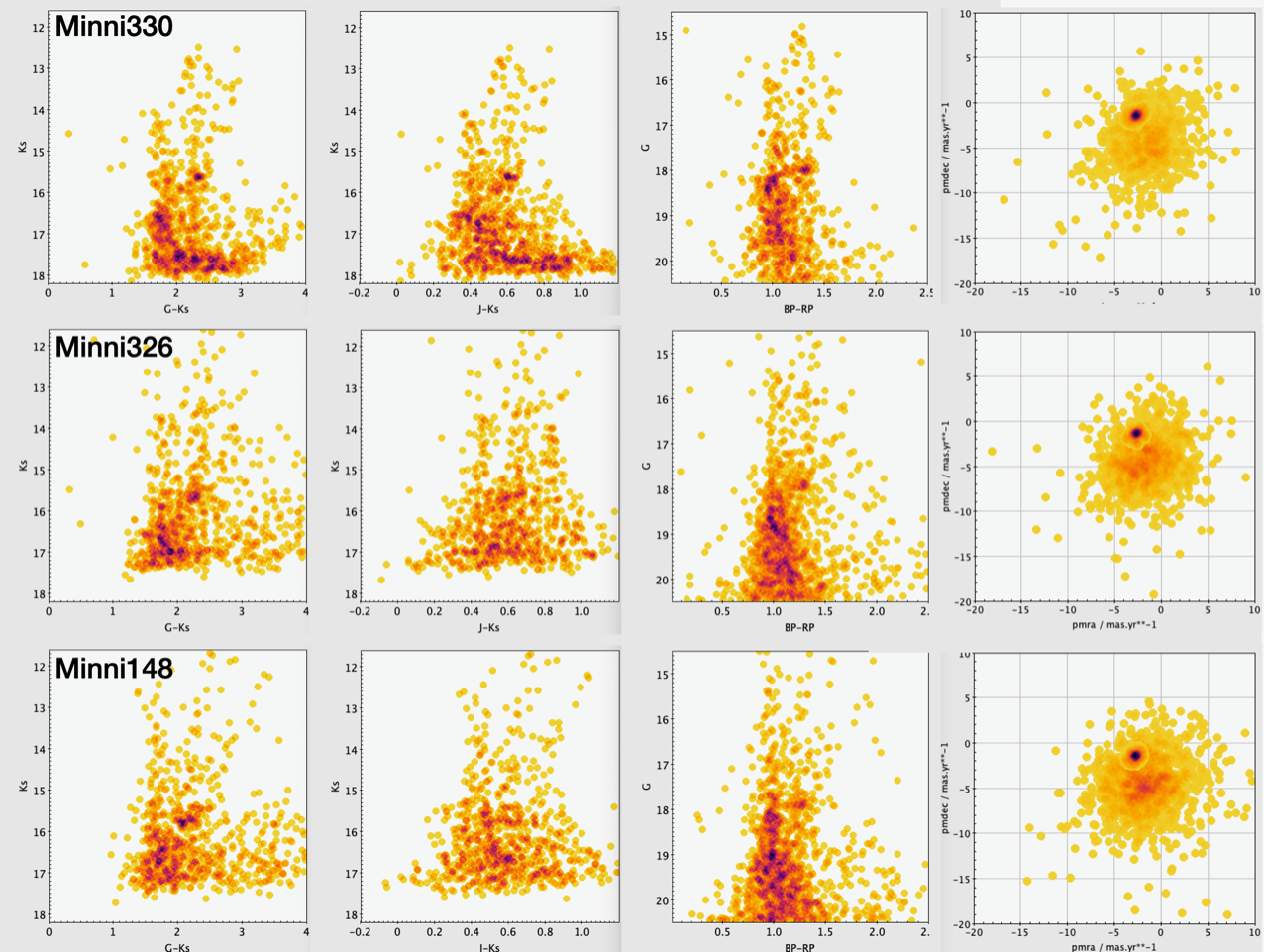} 
\caption{Observed CMDs and vector PM diagrams of the 3’ fields (equivalent to about 23 pc) centered on three
example clusters: Minni148 (bottom), 326 (middle), and 330 (top). 
Even though the RC is conspicuous  in some cases, the GC features appear poorly defined in these CMDs in general because of the overwhelming field contamination.
}
\end{figure*}

The observational data for the candidate GCs and comparison clusters were obtained with the VISTA InfraRed CAMera (VIRCAM) at the 4.1m wide-field Visible and Infrared Survey Telescope for Astronomy (VISTA, see Emerson \& Sutherland 2010) at ESO Paranal Observatory, as part of  the VVV ESO Public Survey (Minniti et al 2010, Saito et al. 2012) and its extension VVVX.
We have been mapping the Galactic bulge and Southern Galactic Plane since 2010, using the J (1.25 $\mu m$), H (1.64 $\mu m$), and $K_s$ (2.14 $\mu m$) near-IR passbands. 
Our data include the main body and northern extension of the Sgr dwarf galaxy (Figure 1).

The deep near-infrared CMDs and LFs are made from point-spread function (PSF) photometry from preliminary VIRAC2 data (Smith et al. 2021, in preparation), an updated version (using a global optimization calibration and secondary distortion correction) of the VVV Infrared Astrometric Catalogue (VIRAC, Smith et al. 2018) for the part mapped by the VVV survey. The part mapped by the VVVX survey uses DoPhot PSF photometry data from Alonso-Garcia et al. (2021, in preparation). 
We also used the $K_s$-band photometry from McDonald et al. (2013, 2014) for only one cluster that is located outside the VVVX area (Minni327).
Even though the GC searches were originally carried out using Gaia-DR2 data (Gaia Collaboration 2018), we only used the more recent Gaia EDR3 photometry and proper motions (PMs) for this work (Gaia Collaboration, Brown et al. 2021).

We initially selected all the cluster candidates observed in the main body of Sgr, within a region of $\sim 180$ sq. deg.
The final list of 23 candidate clusters studied here is given in Table A1.
Figure 1 shows their location in the sky in Galactic coordinates, along with a few previously known GCs in the region.
The areal coverage of the VVV and VVVX near-IR surveys includes the northern half of the main body of Sgr (Figure 1), thus the southern extent remains unexplored.

\begin{figure*}[h]
\centering
\includegraphics[width=10.2cm, height=12.0cm]{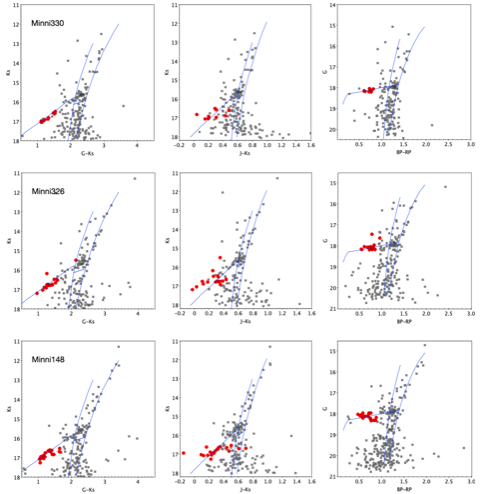} 
\caption{Final PM decontaminated optical and near-IR CMDs within 3' of the RR Lyrae rich GCs Minni148 (bottom), 326 (middle), and 330 (top). 
These decontaminated CMDs show more clearly defined RGB, RC, and HB sequences.
RR Lyrae located within 10' from the cluster centers are also plotted for comparison (red circles).
The thin blue lines depict the mean ridge sequences of the known Sgr GCs NGC 6715 ($[Fe/H]=-1.49$) to the left and Pal 12 ($[Fe/H]=-0.85$) to the right. 
}
\end{figure*}

Five known Sgr GCs are used for comparison: NGC 6715, Arp 2, Ter 7, and Ter 8, which are located in the main body of this galaxy, as well as Pal 12, 
which is located farther away within the Sgr stream (Bellazzini et al. 2020).
In principle, it is safe to assume that these GCs cover comparable reddening and metallicity ranges as the GCs under study, while being located at the same distance ensures that the photometric quality is also similar.
These GCs are  useful to compare the main GC features, as their red giant branch (RGB), red giant branch bump (RGBB), RC, and BHB are tightly defined.

We are searching for GCs in the main body of the Sgr dwarf. These clusters would be located at the Sgr distance of $D=26.5$ kpc (Monaco et al. 2004, Vasiliev \& Belokurov 2020) and share the Sgr kinematics.
Along the line of sight to the Sgr dwarf galaxy, there is contamination from the nearby MW disk and from the more distant bulge field stars. 
Figure 2 shows the observed optical and near-IR CMDs for the 3' fields  centered in the GCs Minni330, 326, and 148, along with their respective Gaia-EDR3 vector PM diagrams. It is clear that the foreground population largely outnumbers the GC stars.

To reduce the amount of contamination in the CMDs,
we first made a cut to discard all stars with parallaxes larger than 0.5 mas, eliminating the nearby disk component. 
{ Then, Sgr GC members were selected by including stars 
within 1.5 mas/yr from the mean Sgr PM vector measured by Helmi et al. (2018), because this value was found adequate for the selection of the Gaia EDR3 PM members of the comparison Sgr GCs, as shown in the appendix.}

The cleaned optical and near-IR CMDs for the GCs Minni330, 326, and 148 are shown in Figure 3, along with the mean ridge lines of a metal-poor and a metal-rich Sgr GC for comparison (NGC 6715 and Pal 12, respectively).
The cluster sequences are now very well defined in all these CMDs, clearly showing the clusters RGB, RC, and BHB. 
The corresponding G- and $K_s$-band  LFs are plotted in Figure 4, and they exhibit a very prominent RC.
The observed RC peaks were used to determine the GC distances, as discussed below in Section 4.

After the foreground contamination was culled out, all the candidate cluster CMDs were compared with the CMDs of known Sgr GCs and with nearby background Sgr fields covering similar area. 
Based on the CMD morphology of these candidate clusters, we have decided that 12 of them are bona fide GCs.
These GCs are listed in Table A1, along with their positions and other physical quantities measured in this work. 
The candidates that are unconfirmed  are labeled as inconclusive, as their CMDs either contain too few stars or they are not very different from the field.

\begin{figure}[h]
\centering
\includegraphics[width=7.0cm, height=6.0cm]{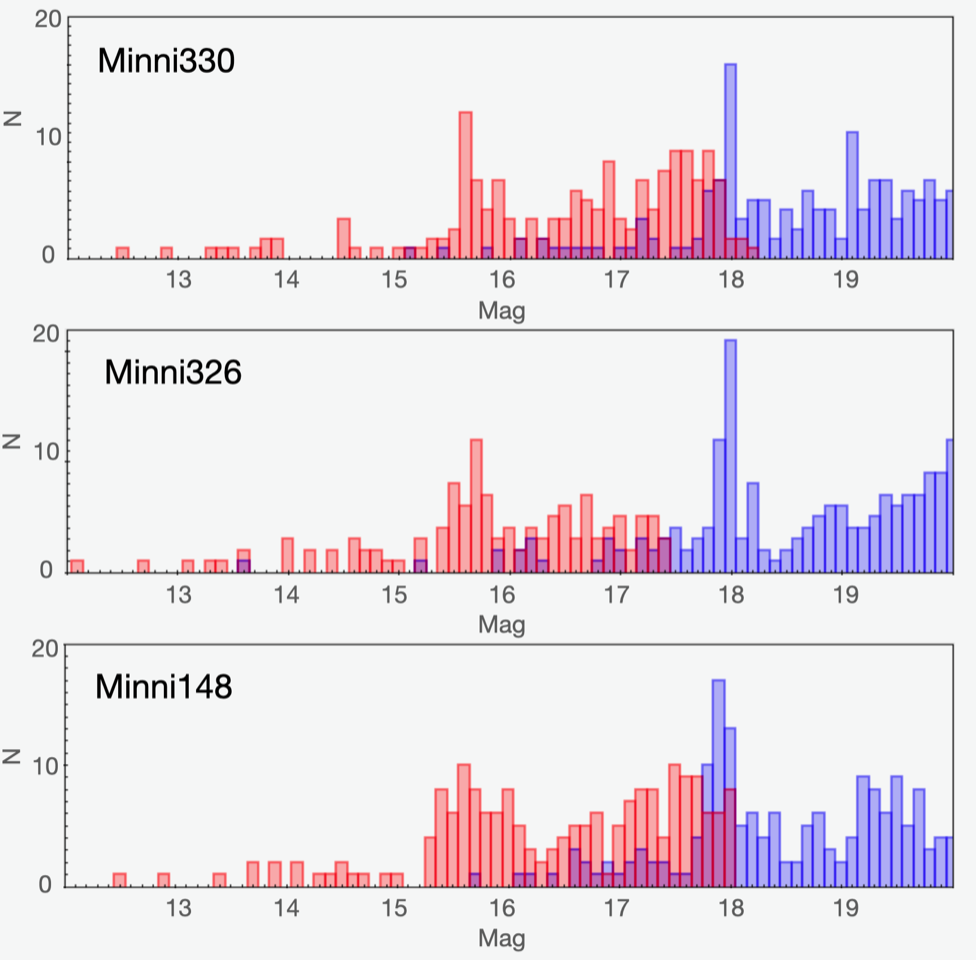} 
\caption{Final PM decontaminated optical and near-IR  LFs within 3' of the GCs Minni148 (bottom), 326 (middle), and 330 (top). 
The $K_s$- and G-band LFs are shown in red and blue, respectively.
We note that the RC peak of the GC is clearly seen in these LFs.  
}
\end{figure}

\section{RR Lyrae}

The Sgr dwarf galaxy is known to contain a rich RR Lyrae population 
(e.g., Alcock et al. 1997, Newberg et al. 2003, Soszynski et al. 2014, Sesar et al. 2017, Ramos et al. 2020).
We have also searched for all RR Lyrae known within 10' fields centered on the target GCs (equivalent to 77 pc at the distance of Sgr), for which Gaia-EDR3 data are available. This search includes RR Lyrae from the samples of MACHO (Alcock et al. 1997), Ogle (Soszynski et al. 2014), Gaia-DR2 (Clementini et al. 2019), and VVV (Majaess et al. 2018). 
After discarding bulge RR Lyrae using the Gaia-EDR3 vector PM diagrams, the numbers of RR Lyrae were found within 3' and 10' from the GC centers they are listed in Table A1.

We find that 21 out of 23 GC candidates in Table A1 contain RR Lyrae within 10' fields, the exceptions being Minni146 and Minni03, which contain no RR Lyrae with the correct PMs and distances in order to belong to the Sgr galaxy.
Also, there are ten objects that do not have any Sgr RR Lyrae within 3' (Minni02, 03, 08, 146, 310, 311, 312, 313,  327, and 329). 
In summary, there are 13 GC candidates with at least one RR Lyrae within 3', with the correct PM and a distance of Sgr.

We measured a mean density of field RR Lyrae belonging to the Sgr dwarf galaxy of 40 RRL/sq. deg. in these fields. 
This allowed us to predict that there should be about 0.3 RR Lyrae in each field of 3’ radius centered on the GCs on average. 
Therefore, GCs that contain just two or three RR Lyrae within 3’ represent a significant overdensity, albeit with small number statistics.
Indeed, six clusters from Table A1 exhibit such an RR Lyrae excess: Minni87, 145, 148, 324, 325, and 326. 
In addition, Minni328 and 330 have only one RR Lyrae within 3’, but many more within 10’ (nine and 11 in total, respectively, where only three are expected), indicating an extended excess of RR Lyrae.
We note that we only counted objects with PMs and magnitudes consistent with Sgr dwarf membership, excluding the brighter foreground RR Lyrae that belong to the Galactic bulge. 

Based on the CMDs and the RR Lyrae, we conclude that these objects are convincing GC members of the Sgr galaxy. 
We also include among the bona fide GCs, the clusters Minni144, 146, 147, 311, and 329, based on the morphology of their optical and near-IR CMDs, even though they contain fewer RR Lyrae.

\section{Distances and other physical parameters}

The most important parameter to measure initially is the distance in order to firmly establish the membership of the new  GCs to the Sgr dwarf galaxy.
The distances were measured using the magnitude of the RC and the period-luminosity relation for RR Lyrae.
We note that the RGBB is prominent in the Sgr dwarf (Monaco et al 2002), but for the distance measurement we used the RC peak in the LF that is due to the combined contribution of the RGBB and red core-He burning  HB stars. Contamination by the RGBB, being fainter than the RC proper, would result in systematically larger distances.

In order to apply the appropriate reddening corrections, we used the maps of Schlafly \& Finkbeiner (2011),
adopting the following relations for the extinctions and reddenings: 
$A_{Ks}=0.11 A_V$, 
$A_{Ks}=0.72 E(J-Ks)$, 
$A_G=0.79 A_V$, and
$A_G=2.0 E(BP-RP)$.
The cluster fields exhibit a narrow range of extinctions and reddenings: 
$0.04<A_{Ks}<0.14$ mag and $0.06<E(J-Ks)<0.19$ mag. Therefore, in the cases of these Sgr GCs, reddening is fairly uniform and not a problem, contrary to the bulge GCs where the large and variable extinction poses a significant challenge (Minniti et al. 2017).

After accounting for the individual reddenings, the RC distances were computed assuming RC absolute magnitudes of
$M_{Ks}=-1.606 \pm 0.009$ and $M_G = (0.495 \pm 0.009) + (1.121 \pm 0.128) (G - Ks - 2.1)$ mag from Gaia DR2  (Ruiz-Dern et al. 2018).
The RC distances $D_{RC}$ measured for the individual candidate clusters are listed in Table A1. 

We also measured individual distances to the RR Lyrae in the cluster fields, after appropriately correcting for reddening.
We used the period-luminosity relation of Muraveva et al. (2015), applied to the fundamental pulsators (RRab) and also to the first-overtone pulsators (RRc), after fundamentalizing them by adding 0.127 to the logarithm of the period.
The mean distances measured using all the RRab and RRc within 10' of the clusters that share the Sgr PMs are also listed in Table A1.
In general, the RC distances are $\sim 2$ kpc larger in the mean than the RR Lyrae distances, an effect that can be due to the contamination to the RC from the RGBB.
We find that the majority of the new GCs are located at the distance of the Sgr dwarf, with their distance moduli differing from the field Sgr population by less than 0.3 mag.

The objects listed in Table A1 are finally classified as bona fide Sgr GCs if they are located at the correct distance and present typical GC CMDs with relatively narrow RGBs. We confirm 12 Sgr GCs, while
the remaining 11 objects are classified as inconclusive when their CMDs do not have clearly defined GC sequences, containing few PM-selected stars, and/or they are not sufficiently different from the background.

This new sample of Sgr GCs needs to be followed up on in order to measure their physical parameters (metallicities, luminosities, and other structural and orbital parameters).
Even though we did not measure these parameters here because they require a more laborious procedure, which will be presented in a follow-up paper, we can conclude that they are low luminosity GCs, clearly fainter than the known Sgr GCs used for comparison here.
{ We note that the addition of these faint GCs would imply that the Sgr GC system deviates from a universal Gaussian shape GC LF as seen in other galaxies (see Rejkuba 2012).
For example, Huxor et al. (2014) have also found an excess of low-luminosity GCs in the M31 GC system.}
Regarding the metallicities, we can also conclude, based on the comparison of the CMDs with those of known GCs, that they seem to be consistent with Sgr GC metallicities.

Finally, Figure 1 clearly shows that there is a vast region of the Sgr galaxy that remains unexplored.  
There were nine GCs previously known in this galaxy, plus 12 new ones confirmed from this work. 
Assuming symmetry, there could be a similar number of GCs in the southern extension.
Hence, using a filling factor of about 40\% for the fields not covered by the VVV/VVVX surveys, and without counting 11 unconfirmed candidates, we estimate the total number of Sgr GCs to be at least 30. 
We note that the vast majority of these new members would also be low luminosity GCs.

\section{Conclusions}
We have identified 23 candidate GCs that may belong to the Sgr dwarf galaxy from the VVV/VVVX surveys searches.
We studied the PM selected optical and near-IR CMDs, and also the RR Lyrae in the cluster fields, measuring their distances.
Weighting all the evidence, we report the discovery of 12 new low luminosity GCs in the main body of the Sgr dwarf, more than doubling the list of confirmed GC members of this nearby galaxy that is an important on-going merger with the MW. 
Since our near-IR surveys do not cover the southern extension of the Sgr galaxy, we predict that several more GCs have yet to be found, estimating a total population of at least 30 GCs in this galaxy.
The predictions of recent theoretical studies of the GC populations in dwarf galaxies (e.g., Kruijssen et al. 2019, 2020, El Badry et al. 2019, Choski \& Gnedin 2019, Hughes et al. 2019) may be compared with the GCs of Sgr once the metallicities, luminosities, and ages of the newly discovered GCs are properly measured (E. Garro et al. 2021, in preparation).
\\

\begin{acknowledgements}
We gratefully acknowledge the use of data from the ESO Public Survey program IDs 179.B-2002 and 198.B-2004 taken with the VISTA telescope and data products from the Cambridge Astronomical Survey Unit. 
D.M. and M.G. are supported by Fondecyt Regular 1170121 and by the BASAL Center for Astrophysics and Associated Technologies (CATA) through grant AFB 170002. 
J.A.-G. acknowledges support from Fondecyt Regular 1201490 and from ANID's Millennium Science Initiative ICN12\_009, awarded
to the Millennium Institute of Astrophysics (MAS). R.K.S. acknowledges support from CNPq/Brazil through project 305902/2019-9.
\end{acknowledgements}

\bibliographystyle{aa.bst}
\bibliography{bibliopaper}

{ References}
\\

Alcock, C., Allsman, R. A., Alves, D. R.,  et al. 1997, ApJ, 474, 217A

Antoja, T., Ramos, P., Mateu, C., et al. 2020, A\&A, 635, L3

Arca-Sedda, M., \& Capuzzo-Dolcetta, R. 2014, ApJ, 785, 51

Barba, R., Minniti, D., Geisler, D., et al. 2019, ApJL, 870, L24

Beasley, M.A., Leaman, R., Gallart, C., et al. 2019, MNRAS, 487, 1986B

Bellazzini, M., Ibata, R., Malhan, K., et al. 2020, A\&A, 636, A107

Benjamin, R. A., Churchwell, E., Babler, B. L., et al. 2003, PASP, 115, 953B

Belokurov, V., Erkal, D., Evans, N. W., Koposov, S. E., \& Deason, A. J. 2018, MNRAS, 478, 611

Belokurov, V., Zucker, D. B., Evans, N. W., et al. 2006, ApJ, 642, L137

Borissova, J., Chené, A.-N., Ramírez Alegría, S., et al., 2014, A\&A, 569, A24

Burkert, A., \& Forbes, D. 2020, AJ, 159, 56

Camargo, D. \& Minniti, D. 2019, MNRAS, 484, L90C

Choksi, N., \& Gnedin, O. Y. 2019, MNRAS, 488, 5409 

Clementini, G., Ripepi, V., Molinaro, R., et al. 2019, A\&A, 622, A60

Correnti, M., Bellazzini, M., Ibata, R. A., Ferraro, F. R., \& Varghese, A. 2010, ApJ, 721, 329

Crnojevic, D., Sand, D. J., Zaritsky, D., et al. 2016, ApJ, 824, L14

Cutri, R. M.; et al. 2012, VizieR On-line Data Catalog: II/311

El Badry, K., Quataert, E., Weisz, D. R., et al.  2019, MNRAS, 482, 4528

Emerson, J., \& Sutherland, W. 2010, The Messenger, 139, 2E

Fernández-Trincado, J. G.,  Beers, T. C.,  Minniti, D., et al. 2020b, A\&A Letters in press (arXiv:2010.01135)

Fernández-Trincado, J. G.,  Beers, T. C.,  Placco, V. M., et al. 2020a, ApJ, , 903, L17F

Gaia Collaboration, Brown, A. G. A., Vallenari, A., et al. 2018, A\&A, 616, A1 

Gaia Collaboration, Brown, A. G. A., Vallenari, A., et al. 2021, A\&A, in press (arXiv:2012.01533)

Gaia Collaboration, Helmi, A., van Leeuwen, F., et al. 2018, A\&A, 616, A12

Gnedin, N.Y., \& Ostriker, J.P, 1997, ApJ, 486, 581G

Goerdt, T., Moore, B., Read, J. I., et al. 2006, MNRAS, 368, 1073

Hasselquist, S., Carlin, J. L., Holtzman, J. A., et al. 2019, ApJ, 872, 58
 
Hasselquist, S., Shetrone, M., Smith, V., et al. 2017, ApJ, 845, 162

Hayes, C. R., Majewski, S. R., Hasselquist, S., et al. 2020, ApJ, 889, 63

Helmi, A., Babusiaux, C., Koppelman, H. H., et al. 2018, Nature, 563, 85

Huang, K.-W. \& Koposov, S. E. 2020, MNRAS in press (arXiv:2005.14014)

Hughes, M. E., Pfeffer, J., Martig, M., et al.  2019 MNRAS, 482, 2795

Huxor, A. P., Mackey, A. D., Ferguson, A. M. N., et al. 2014, MNRAS, 442, 2165

Ibata, R. A., Gilmore, G., \& Irwin, M. J. 1994, Nature, 370, 194

Kruijssen, J. M., Pfeffer, J. L., Reina-Campos, M.,  et al. 2019, MNRAS, 486, 3180 

Kruijssen, J. M., Pfeffer, J. L., Chevance, M.,  et al. 2020, MNRAS, 498, 2472

Laporte, C. F., Johnston K. V., Gomez, F., Garavito-Camargo, N., \& Besla, G., 2018, MNRAS, 481, 286

Law D., \& Majewski S. 2010, ApJ, 718, 1128

Lotz, J. M., Telford, R., Ferguson, et al. 2001, ApJ, 552, 572

Majaess, D., Dekany, I., Hagdu, G., et al. 2018, ApSS, 363, 127M

Majewski, S. R., Skrutskie, M. F., Weinberg, M. D., \& Ostheimer, J. C. 2003, ApJ, 599, 1082

Majewski, S.R., Kunkel, W., Law, D.R. et al. 2004, AH, 128, 245M

Massari, D., Koppelman, H. H., \& Helmi, A. 2019, A\&A, 630, L4

Mateo, M., Mirabal, N., Udalski, A., et al. 1996, ApJ, 458, L13

McDonald, I., Zijlstra, A. A., Sloan, G. C., et al., 2013, MNRAS, 436, 413 

McDonald, I., Zijlstra, A. A., Sloan, G. C., et al., 2014, MNRAS, 439, 2618M

Minniti, D., Fernández-Trincado, J. G., Ripepi, V., et al., 2018, ApJ, 869, L10

Minniti, D., Gomez, M., Pullen, J. B., et al. 2020, RNAAS, 4, 218M

Minniti, D., Hempel, M., Toledo, I., et al. 2011, A\&A, 527, A81

Minniti, D., Geisler, D., Alonso Garcia, J., et al., 2017, ApJ, 849, L24

Minniti, D., Lucas, P. W., Emerson, J. P., et al. 2010, NewA, 15, 433 

Monaco, L., Bellazzini, M., Ferraro, F. R., \& Pancino, E. 2002, ApJL, 578, 47M

Monaco, L., Ferraro, F. R., Bellazzini, M., \& Pancino, E. 2004, MNRAS, 353, 874M

Moni Bidin, C., Mauro, F., Geisler, D.,  2011, A\&A, 535, A33

Montenegro, K., Minniti, D., Alonso-García, J., et al. 2019, ApJ, 872, 206M

Muraveva, T., Palmer, N., Clementini, G., et al. 2015, MNRAS, 446, 3034M

Myeong, G.C., Vasiliev, E., Lorio, G., et al. 2019, MNRAS, 488, 1235M

Newberg, H. J., Yanny, B., Cole, N., et al. 2007, ApJ, 668, 221

Newberg, H. J., Yanny, B., Grebel, E. K., et al. 2003, ApJ, 596, L191

Palma T., Minniti, D., Alonso-Garcia, J., et al., 2019, MNRAS, 487, 3140

Ramos, P., Mateu, C., Antoja, T., et al. 2020, A\&A, 638, A104

Rejkuba, M. 2012, ApSS, 341, 195

Ruiz-Dern, L., Babusiaux, C., Arenou, F., Turon, C., \& Lallement, R. 2018, A\&A, 609, A116

Saito, R. K., Hempel, M., Minniti, D., et al. 2012, A\&A, 537, A107

Schlafly, E. F., \& Finkbeiner, D. P., 2011, ApJ, 737, 103

Schlafly, E. F., Green, G. M., Lang, D., et al. 2018, ApJS, 234, 39S

Sesar, B., Hernitschek, N., Dierickx, M. I. P., Fardal, M. A., \& Rix, H.-W. 2017, ApJ, 844, L4

Skrutskie, M. F., Cutri, R. M., Stiening, R., et al. 2006, AJ, 131, 1163

Smith, L. C., Lucas, P. W., Kurtev, R., et al. 2018, MNRAS, 474, 1826

Soszynski, I., Udalski, A., Szymanski, M.K., et al., Acta Astron. 2014, 64, 177S

Tremaine S.,  \& Weinberg M. D. 1984 MNRAS 209 729

Vasiliev, E., 2019, MNRAS, 484, 2832

Vasiliev, E.,  \& Belokurov, V., 2020, MNRAS 497, 4162

Vasiliev, E., Belokurov, V., \& Erkal, D. 2021,  MNRAS, in press (arXiv:2009.10726)

Wang, M. Y., Koposov, S., Drlica-Wagner, A., et al. 2019, ApJ, 875L, 13W

\begin{appendix}
\section{CMDs, LFs, and finding charts}

Table A1 below lists the data for all the GC candidates analyzed. We give the 
IDs, 
positions in equatorial coordinates (J2000), 
near-IR and optical extinctions $A_{Ks}$ and $A_G$, 
distance measured using the RC, 
mean near-IR and optical RC magnitudes ($Ks_{RC}$ and $G_{RC}$), 
number of RR Lyrae within 3' and 10' from the cluster centers (N3' and N10'), 
distance measured using the RR Lyrae, 
and survey data and 
our final classification as either a confirmed  GC located at the distance of the Sgr dwarf galaxy or an inconclusive GC candidate requiring additional data for confirmation or rejection. 

The optical and near-IR CMDs and G- and Ks-band LFs for all these candidates are shown in Figures 1 to 4 of this appendix.
The Gaia EDR3 vector PM diagrams and optical CMDs of the previously known Sgr GCs used for comparison are shown in Figure 5.
Finally, Figure 6 shows the 3' $\times$ 3' finding charts for these targets, which are severely outnumbered by the foreground stars along the line of sight.

\end{appendix}

\begin{table}
\onecolumn
\centering 
\caption{GC candidates in the Sgr galaxy.}
\begin{tabular}{lcllllllrrllllll}
\hline\hline
ID             &RA      & DEC  & $A_{Ks}$ &$A_G$   & $D_{RC}$  & $Ks_{RC}$ & $G_{RC}$  & N3'  & N10' &$D_{RRL}$ &  Survey & Classification \\          
    & (J2000) & (J2000)&[mag]&[mag] &[kpc] &[mag]&[mag]& & &[kpc]&  & \\
\hline
Minni01   & 18 34 48 & -28 42 40   & 0.08& 0.55 &$26.7\pm 0.6$ & 15.60 & 18.15 &    1 &   3 &  $26.1\pm 0.7$    & VVV       & inconclusive  \\
Minni02   & 18 30 02 & -28 26 24   & 0.11& 0.83 &…                     & …       &  …    &    0 &   2 &  $25.6\pm 0.7$    &  VVV       & inconclusive  \\
Minni03  & 18 20 23 & -32 24 29   & 0.06& 0.45 &…                      & …       &  …    &    0 &   0 &  …                       &   VVV      & inconclusive  \\
Minni04  & 18 15 35 & -28 18 00   & 0.11& 0.79 &…                      & …       &  …     &   1 &   2 &  $23.0\pm 2.0$    &   VVV      & inconclusive  \\
Minni08   & 18 22 19 & -26 37 41  & 0.14& 0.98 &…                      & …       &  …    &    0 &   3 &  $26.2\pm 0.6$    &  VVV       & inconclusive  \\
Minni84   & 18 22 24 & -30 39 09  & 0.08& 0.55 &…                      & …       &  …    &    1 &   5 &  $25.7\pm 0.9$    &  VVV       & inconclusive   \\
Minni87  & 18 13 48 & -27 57 23   & 0.14&1.01 &…                       & …       & 17.9: &   2 &   3  &  …                       &   VVV      & inconclusive \\
Minni144 & 18 40 09 & -30 34 26  & 0.05& 0.32 &$28.3\pm 0.6$  &  15.7:  & 17.95 &   1 &   7 &  $25.4\pm 1.0$   & VVVX,MD& Sgr GC  \\
Minni145 & 18 53 28 & -28 33 44  & 0.06& 0.40 &$28.8:$             & …       & 18.10 &   5 &  14 &  $27.2\pm 2.8$   & VVVX,MD& Sgr GC  \\
Minni146 & 18 52 12 & -31 56 45  & 0.04& 0.28 &$27.5\pm 0.6$  & 15.60 & 18.00  &   0 &   0 &  …                       & VVVX,MD& Sgr GC \\
Minni147 & 18 41 56 & -30 56 07  & 0.04& 0.27 &$27.8\pm 0.6$  & 15.70 & 18.05  &   1 &   6 &  $25.9\pm 0.9$   & VVVX,MD& Sgr GC \\
Minni148 & 18 51 36 & -30 24 58  & 0.04& 0.31 &$27.7\pm 0.6$  &  15.65 & 17.90 &  3 &  20 &  $25.7\pm 1.4$   & VVVX,MD& Sgr GC \\
Minni310 & 18 34 15 & -28 33 49  & 0.08& 0.57 &$29.3:$             & 15.8:   & 18.2:  &   0 &   1 &  …                      &  VVV        & inconclusive   \\
Minni311 & 18 34 30 & -28 47 03  & 0.08& 0.58 &$30.6:$             & 15.9:   & 18.3:  &   0 &   3 &  $26.5\pm 0.2$   &  VVV        & Sgr GC   \\
Minni312 & 18 34 38 & -28 51 37  & 0.08& 0.60 &$29.1:$             &  15.8:  & 18.1: &    0 &   2 &  $26.1\pm 0.5$   &   VVV       & inconclusive\\
Minni313  & 18 45 33 & -24 09 08  & 0.11 & 0.80 &…                     & …     &  …     &    0 &    3 &  …                      &  VVV         & inconclusive   \\
Minni324  & 18 46 19 & -29 27 00  & 0.05& 0.38 &$28.8\pm 0.6$  & 15.75 & 17.90 &  2 &  9   &  $25.5\pm 2.0$   & VVVX,MD & Sgr GC \\
Minni325  & 18 54 10 & -31 58 12  & 0.04& 0.26 &$27.8\pm 0.6$  & 15.65 & 17.85 &  0 &  3   &  …                      & VVVX,MD & Sgr GC \\
Minni326  & 18 51 07 & -29 56 24  & 0.05& 0.35 &$28.3\pm 0.6$  & 15.70 & 18.00 &  3 & 17  &  $26.0\pm 2.0$   & VVVX,MD & Sgr GC \\
Minni327  & 19 00 43 & -29 59 24  & 0.04& 0.32 &$27.0\pm 0.6$  &  15.6: &  17.9: &  0 &   5  &  $24.0\pm 3.9$   &  MD           & inconclusive  \\
Minni328  & 18 47 17 & -30 06 00  & 0.05& 0.38 &$26.9\pm 0.6$  & 15.60 & 17.95 &  1 &   9  &  $25.1\pm 1.0$   & VVVX,MD & Sgr GC  \\
Minni329  & 18 44 31 & -30 01 12  & 0.04& 0.29 &$28.5\pm 0.6$  & 15.70 & 17.90 &   0 &  4  &  $24.9\pm 0.7$   & VVVX,MD & Sgr GC \\
Minni330  & 18 55 38 & -29 59 24  & 0.04& 0.31 &$27.7\pm 0.6$  & 15.65 & 18.00 & 1 & 11 &  $26.7\pm 2.0$   & VVVX,MD   & Sgr GC  \\
\hline\hline
\end{tabular}
\label{rrlpat}
\end{table}

\begin{figure*}[h]
\centering
\includegraphics[width=18.2cm, height=17.2cm]{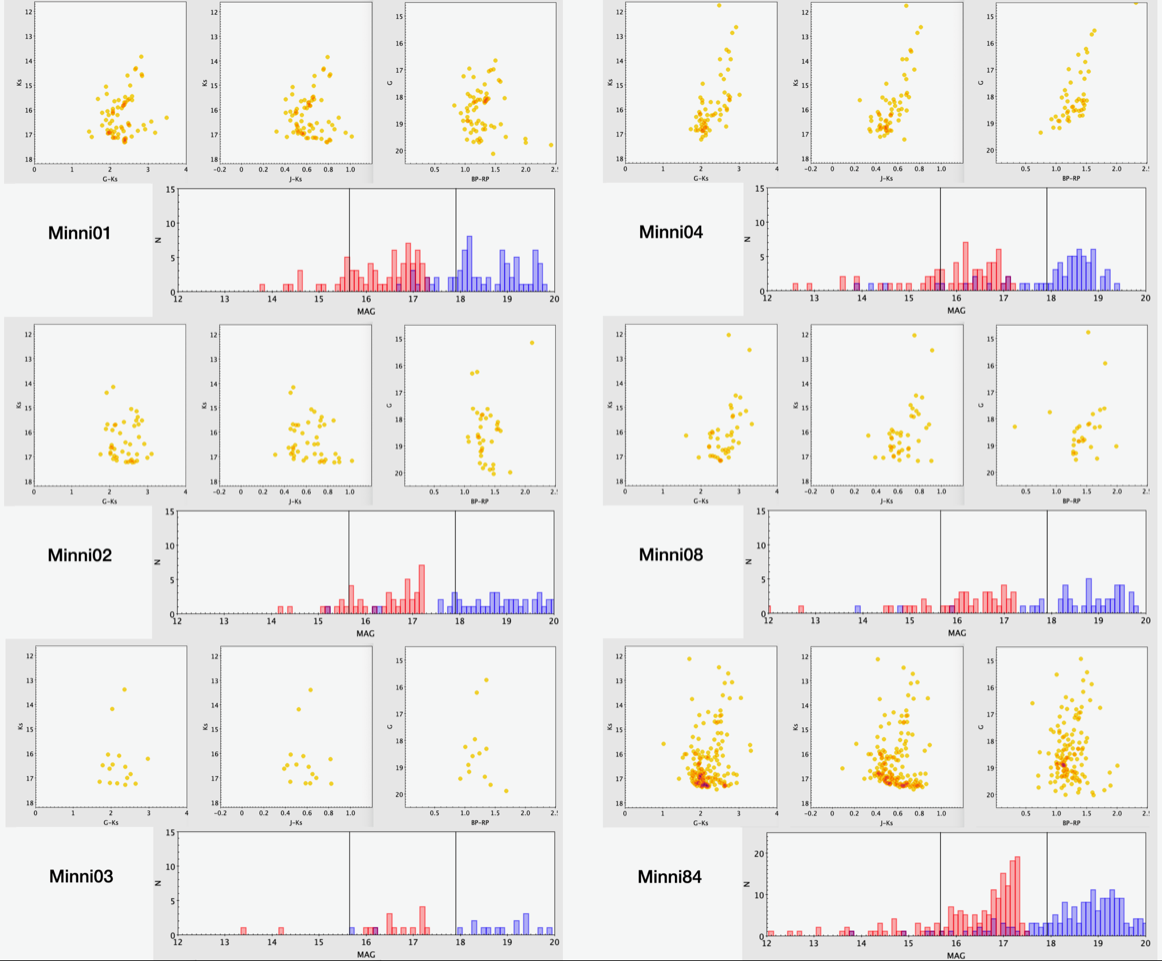} 
\caption{Final PM decontaminated optical and near-IR CMDs within 3' of the GC candidates Minni 01, 02, 03, 04, 08, and 84, along with their G- and Ks-band LFs (blue and red, respectively). 
The last panel shows these diagrams for the comparison GC NGC6715.
The vertical black lines in all the LFs indicate the position of the G- and Ks-band RC peaks for NGC6715.  
}
\end{figure*}

\begin{figure*}[h]
\centering
\includegraphics[width=18.2cm, height=17.2cm]{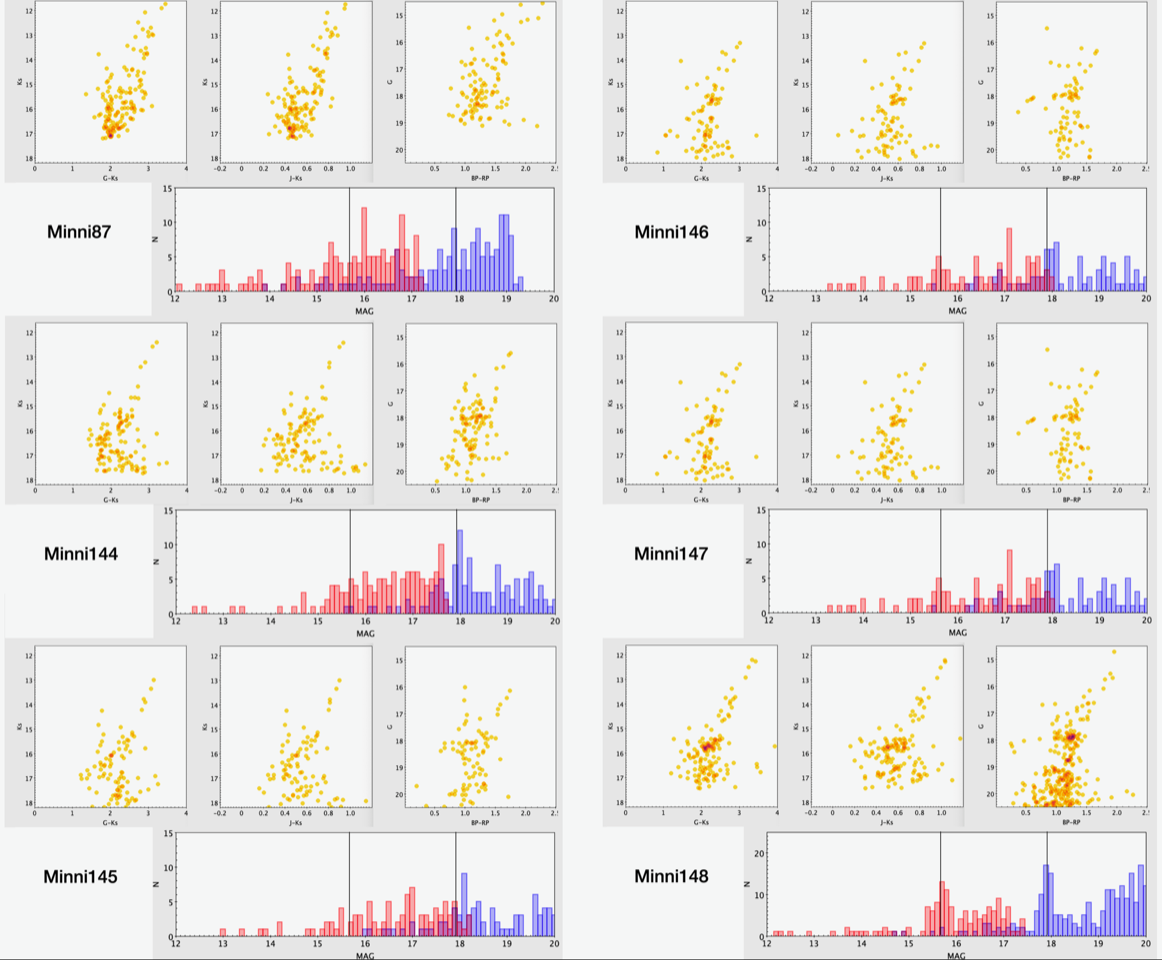} 
\caption{Same as the previous figure, but for the  GC candidates Minni 87, 144, 145, 146, 147, and 148.
}
\end{figure*}

\begin{figure*}[h]
\centering
\includegraphics[width=18.2cm, height=17.2cm]{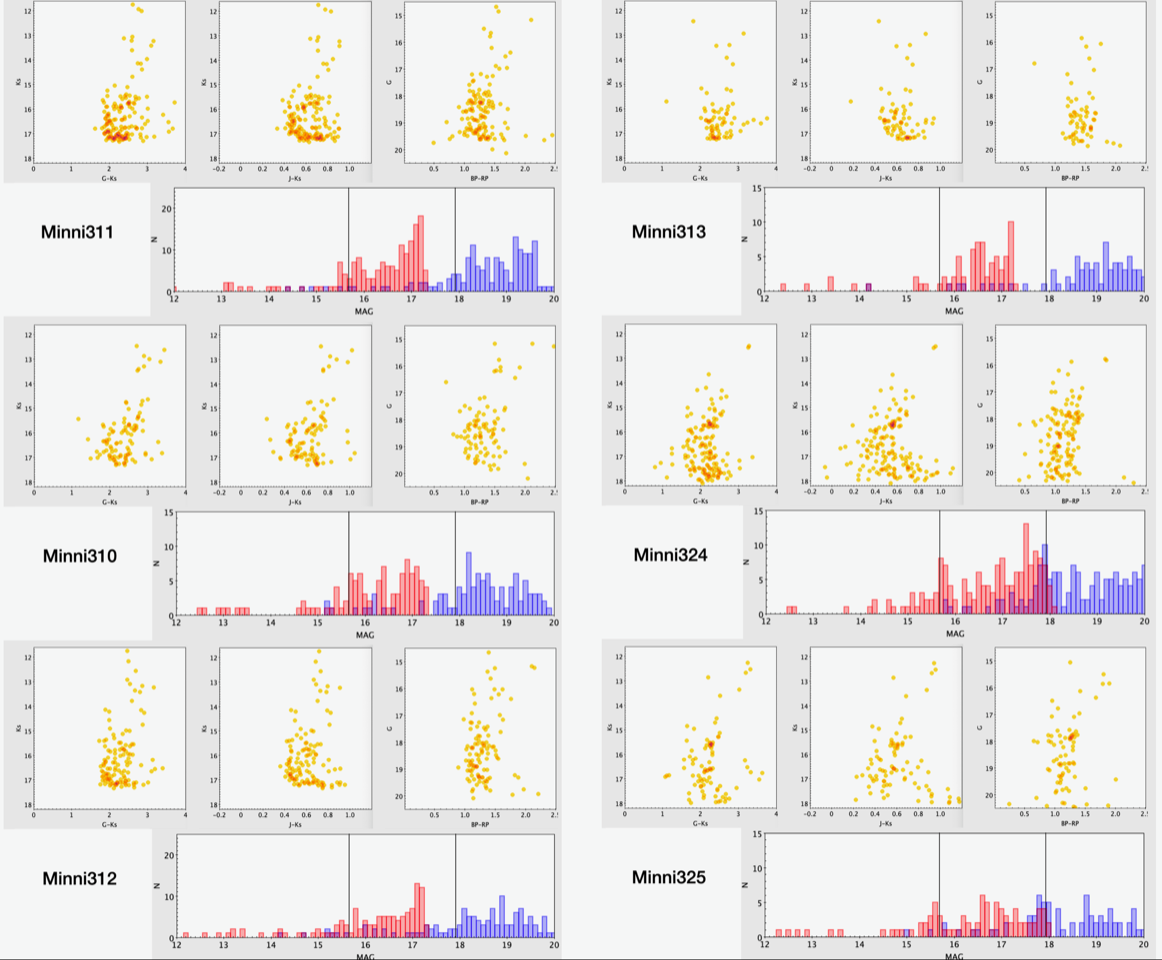} 
\caption{Same as the previous figure, but for the  GC candidates Minni 310, 311, 312, 313, 324, and 325.
}
\end{figure*}

\begin{figure*}[h]
\centering
\includegraphics[width=18.2cm, height=17.2cm]{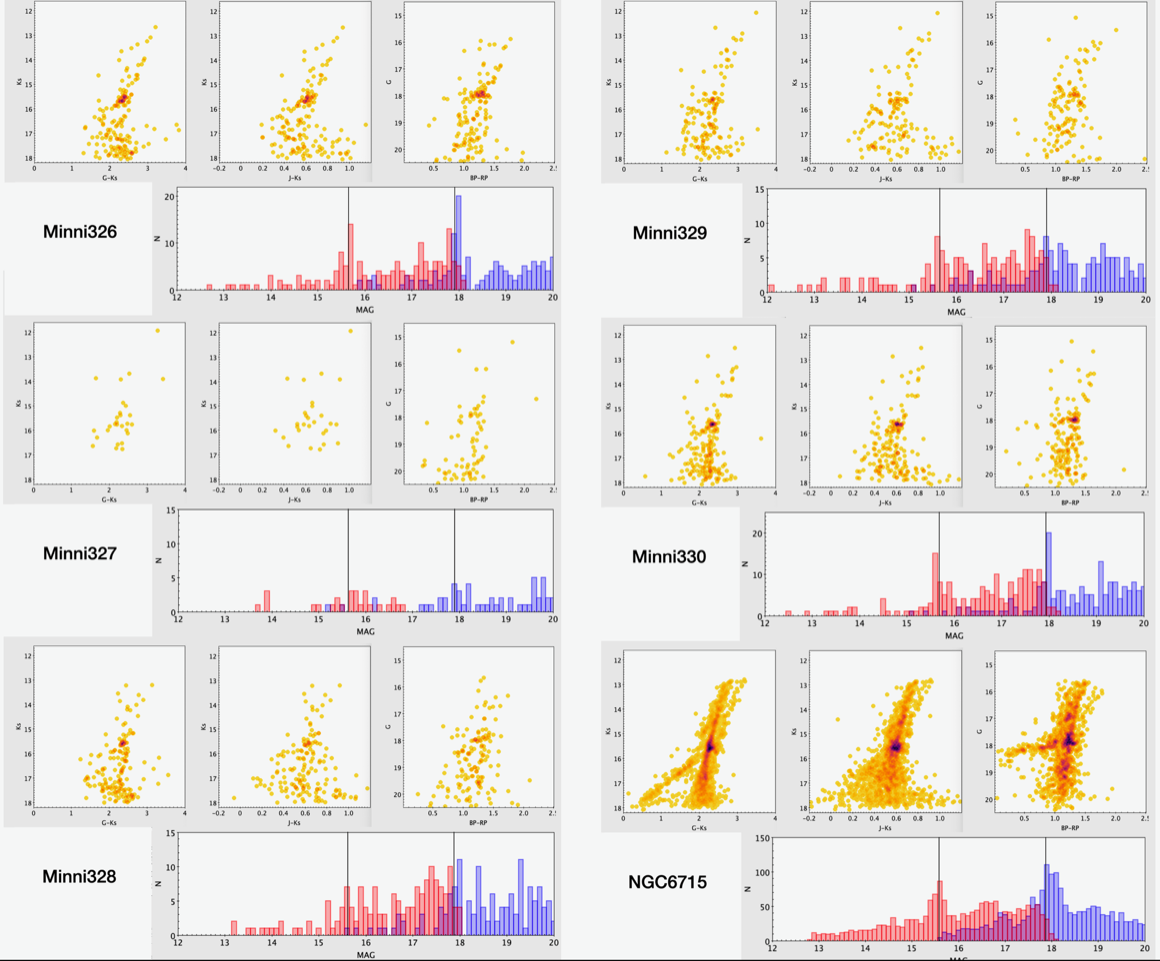} 
\caption{Same as the previous figure, but for the  GC candidates Minni 326, 327, 328, 329, and 330, and the comparison GC NGC 6715.
}
\end{figure*}

\begin{figure*}[h]
\centering
\includegraphics[width=5.2cm, height=9.0cm]{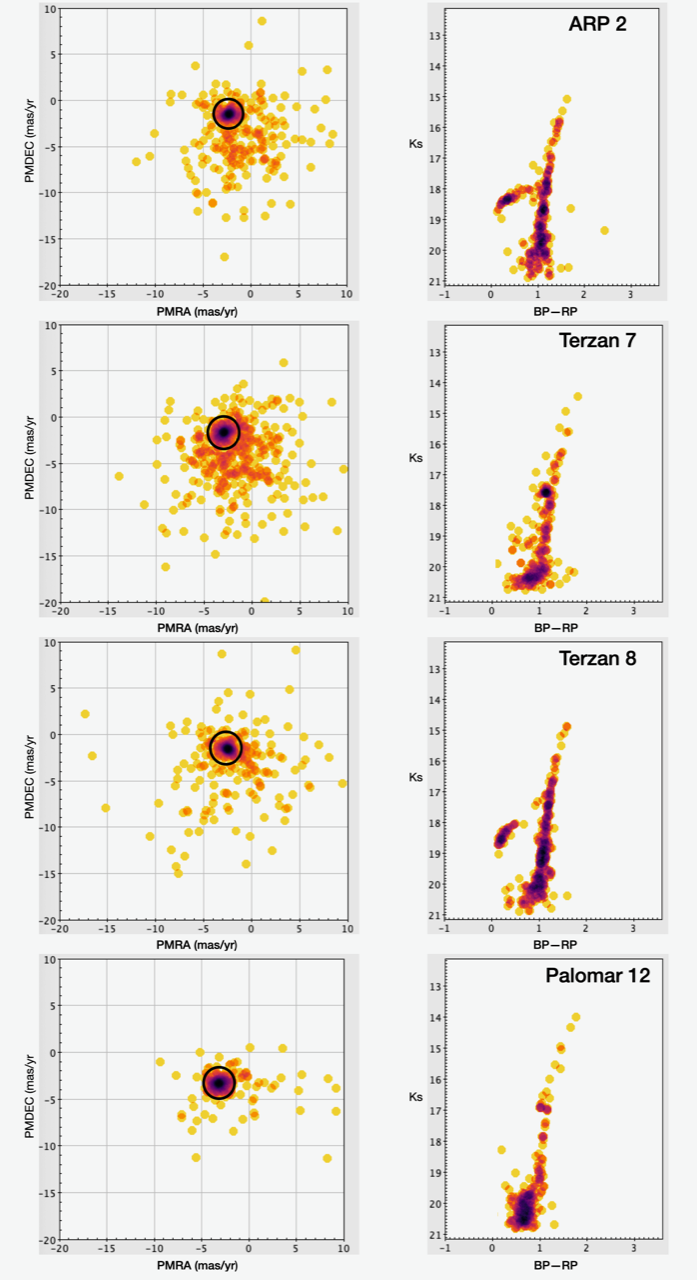} 
\caption{Left panels: Vector PM diagrams for the  known Sgr GCs Arp 2, Ter 7, Ter 8, and Pal 12 used for comparison, with the black circles indicating the selected members.
Right panels: Final PM decontaminated Gaia EDR3 CMDs for the comparison clusters. 
}
\end{figure*}

\begin{figure*}[h]
\centering
\includegraphics[width=12.2cm, height=12.2cm]{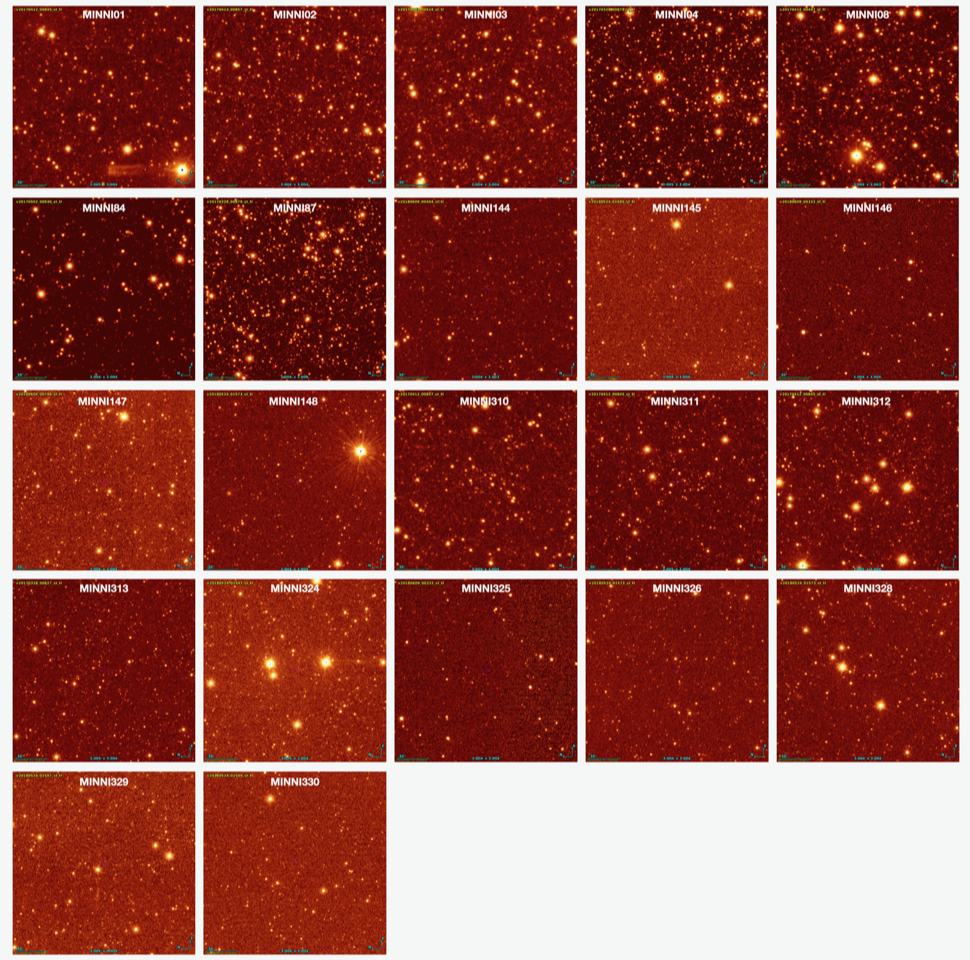} 
\caption{VISTA  3' $\times$ 3' Ks-band images for  the candidate GCs listed in Table 1 (except for Minni 327 which is outside the field of view).}
\end{figure*}

\end{document}